**E-beam manipulation of Si atoms on graphene edges with aberration-corrected STEM**

Notice to the editor (not to be published): This manuscript has been authored by UT-Battelle, LLC, under Contract No. DE-AC05-00OR22725 with the U.S. Department of Energy. The United States Government retains and the publisher, by accepting the article for publication, acknowledges that the United States Government retains a non-exclusive, paid-up, irrevocable, world-wide license to publish or reproduce the published form of this manuscript, or allow others to do so, for United States Government purposes. The Department of Energy will provide public access to these results of federally sponsored research in accordance with the DOE Public Access Plan (http://energy.gov/downloads/doe-public-access-plan).

**E-beam manipulation of Si atoms on graphene edges with aberration-corrected STEM**

*Ondrej Dyck\*, Songkil Kim, Sergei V. Kalinin, Stephen Jesse*

Dr. O. D. Author 1, Dr. S. K. Author 2, Dr. S. V. K. Author 3, Dr. S. J. Author 4
Center for Nanophase Materials Science, Oak Ridge National Laboratory, Oak Ridge, TN, 37830, USA
E-mail: dyckoe@ornl.gov

The burgeoning field of atomic level material control holds great promise for future breakthroughs in quantum and memristive device manufacture and fundamental studies of atomic-scale chemistry. Realization of atom-by atom control of matter represents a complex and ongoing challenge. Here, we explore the feasibility of controllable motion of dopant Si atoms at the edges of graphene via the sub-atomically focused electron beam in a scanning transmission electron microscope (STEM). We demonstrate that the graphene edges can be cleaned of Si atoms and then subsequently replenished from nearby source material. It is also shown how Si edge atoms may be "pushed" from the edge of a small hole into the bulk of the graphene lattice and from the bulk of the lattice back to the edge. This is accomplished through sputtering of the edge of the graphene lattice to bury or uncover Si dopant atoms. These experiments form an initial step toward general atomic scale material control.

1. Introduction

Atomic-scale manufacturing has remained a long-held dream for nanoscience; the ability to specify a material structure at the atomic level and then construct such a material from the atom up. Such capabilities would greatly enhance our understanding of chemistry and physics at the atomic level and, by extension, our understanding of materials, interfaces defects, etc. The first foray into single atom manipulation began in the early 1980s with the work of Eigler at IBM, where he demonstrated the ability of a scanning tunneling microscope (STM) to move single atoms along a surface and construct atomic-scale structures.[1-5] Despite these impressive demonstrations, STM atomic manipulation has several limiting constraints. STM operation requires low temperatures and ultra-high vacuum environments and the atomic structures constructed are typically constrained to reactive surfaces. Given these limitations it seems natural to ask which other techniques might be suited for atomic scale manipulation that may provide a viable pathway around such obstacles.

In recent years, the aberration corrected scanning transmission electron microscope (AC-STEM) has emerged as a powerful imaging and analytic tool for atomic scale studies.[6] These instruments are capable of focusing an electron probe to sub-angstrom dimensions and directing them onto single atoms or atomic columns allowing atomic resolution imaging and spectroscopy. Multiple advances enabled by electron microscopy in understanding materials are summarized in a number of recent books and reviews.[6-8]

It has long been known that electron microscopy techniques can bring about sample damage through electron beam irradiation.[9,10] Historically, changes to a sample brought on by electron beam irradiation in STEM have been considered a detriment, a strike against STEM modalities. Indeed, beam sensitive samples often cannot be investigated with such an instrument because the very act of examining them alters their structure and casts doubt that any observations will be representative of the pristine material. Recently, however, there has been growing interest in attempts to control beam/sample interactions in such a way as bring about intentional atomic scale material modifications[11-18] like vacancy ordering,[16,19-24] single dopant atom motion,[24-27] and chemical reactions.[28,29] Investigations along these lines have begun to appear, with Jesse et. al. publishing a review on atomic scale 3D nanofabrication including STEM based techniques,[30] and Susi et. al. recently publishing an article reviewing the current literature on controllable atomic scale beam/sample interactions in graphene[31].

Nevertheless, the current state of the art of 3D nanofabrication at the atomic level by STEM is but in a nascent state. Myriad sample responses to the electron beam must be thought through, tested, and understood to uncover reproducible methods for atomic scale material control. Here, we discuss the observed behavior of Si atoms attached to graphene edges and present a couple of beam control strategies used to predictably remove or introduce Si atoms onto a graphene edge, move Si atoms along the edge, and move Si atoms from the graphene edge into the lattice as a substitutional defect and back to the edge. While these experiments were performed using Si, they likely extend to other atoms as well, and we discuss the potential technical developments that can accelerate the progress in this area.

## 2. Results and Discussion

We would like to explore controllable beam/sample interactions that may be useful for manipulating matter at the atomic scale. Here we explore the three closely related tasks, namely removing and reintroducing Si edge atoms, moving Si atoms along a graphene edge, and carbon sculpting for the removal or incorporation of dopant atoms. These experiments utilize the capability to irradiate a sample sub-region, realized via creation of a sub-scan box within a reference image. This sub-scan box may be moved with the mouse to direct the beam while concurrently observing the image generated from the sub-scan. We highlight this detail to bring attention to the shortage of available beam control tools. This one rudimentary (but quite useful) tool has made these and previous[32] experiments rather trivial to perform, but without which would be practically prohibitive. Additional discussion regarding beam control tools is presented last.

## 2.1 Removing and Reintroducing Edge Atoms

We first present our experiment on the removal and reintroduction of Si atoms which are often observed passivating the edges of graphene. **Figure 1** a)-d) show the process of removing Si atoms from the edge of a graphene sheet. To accomplish this, a sub-scan area was selected, represented by the dotted box in a). The sub-scanned beam was then moved back and forth over the edge of the hole. This process mobilizes the Si and C atoms at the edge, causing both random reconfigurations and occasional sputtering (i.e. removal of Si or C atoms from the edge).

The combined effect is detailed in b)-d) which show the edge configuration through time. The edge of the graphene lattice which was under e-beam irradiation becomes cleaner until only a single Si atom is left. We note that while Si substitutional atoms are somewhat mobile in the graphene lattice[25, 31-33] we have not observed any beam-induced diffusion from the edge into the bulk of the lattice, even though we attempted to induce this behavior through techniques similar to those used previously[32, 33] to induce directed atomic motion of Si atoms in graphene. In other words, Si atoms are preferentially adsorbed at the graphene edge, which acts as 1D confining potential.

The substitutional Si atoms seen in the initial configuration, a), have not moved further into the lattice. We hypothesize two preventative mechanisms (1) given that the carbon atoms within one or two lattice steps of the edge readily rearrange under the beam[34] and that the carbon atoms adjacent to the Si atoms are less strongly bonded in the lattice,[25] the presence of a Si atom within a few lattice steps of the edge, extends the depth of the rearrangement area. If a Si atom moves toward the edge, the rearrangement area is decreased and the lattice behind the Si atom is healed, which forms a stronger/more robust structure. This results in a higher probability that a Si point defect will randomly migrate toward the edge if it is within a few lattice steps of it. This also explains why we have not often observed Si atoms diffusing from the edge into the graphene lattice.

The second possibility is that (2) the edge is also gradually sputtered away under the 60 kV beam, thus uncovering the slightly buried substitutional defects. While this second mechanism is certainly at play, which we will discuss in the second part, it fails to explain why Si defects have not been observed to move into the lattice from the edge.

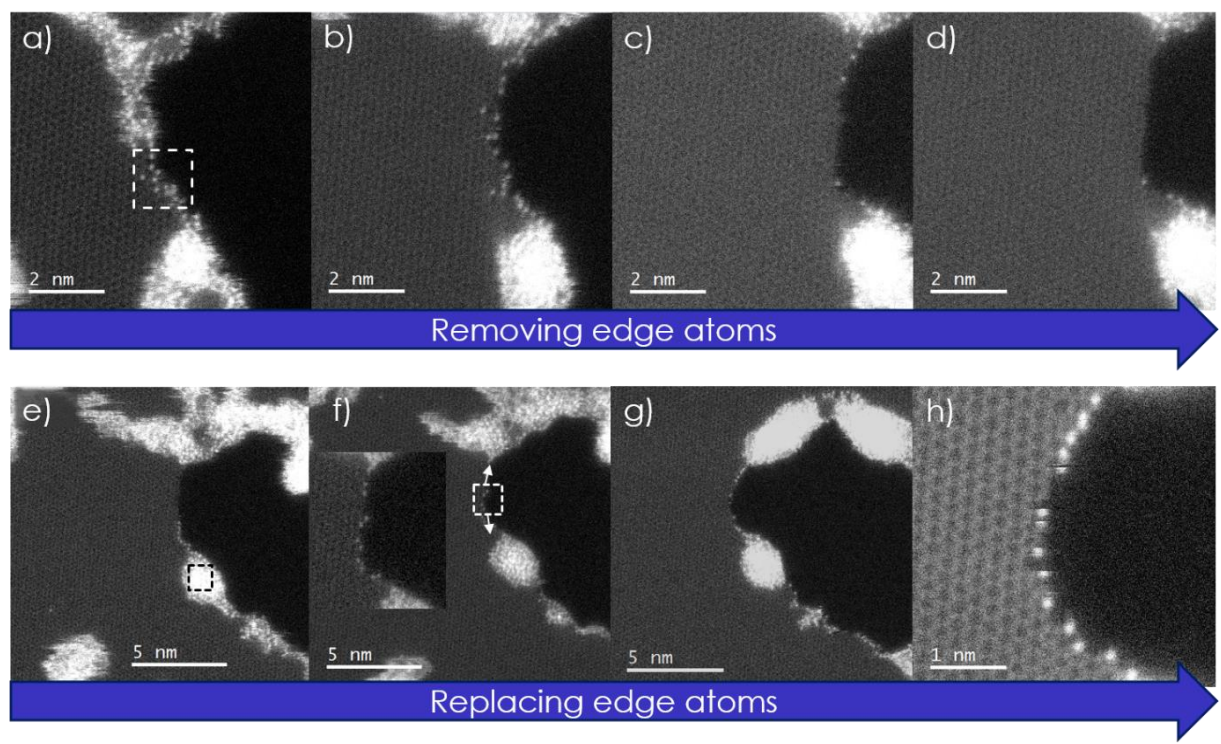

**Figure 1** a)-d) show the process of removing Si edge atoms in a controllable fashion. The dotted box in a) is representative of the sub-scan area used to knock the Si atoms away. This area was moved back and forth, along the edge of the graphene with b)-d) showing the edge evolution through time. After the edge was cleaned of Si atoms the sub-scan was placed over the Si source material represented by the box in e). The Si atoms were sputtered from the source material and populated the attachment sites along the graphene edge for about 2 nm. The sub-scan area was then dragged back and forth along the edge to mobilize the Si edge atoms, illustrated by the box and arrows in f). The inset in f) shows an enlarged view of the edge after this procedure where we can see that the Si atoms have dispersed randomly across the edge. Continued sub-scanning over the source material, both above and below the graphene edge, attaches more Si atoms to the edge, shown in g). Finally, h) shows a magnified view of the edge in the final state. We note that Si atoms are removed by the imaging process itself so it is difficult to fully populate the edge attachment sites while also detecting that they are indeed filled.

Figure 1 e)-h) detail the process used to "push" Si atoms from the nearby source material back onto the cleaned edge. A sub-scan area was selected, indicated by the dotted box in e), and the beam was scanned over just the source material, sputtering atoms away. As can be observed in e), this procedure resulted in every Si attachment site being occupied by a Si atom within about 2 nm from the source material. Continuing this procedure did not noticeably increase the number of Si atoms attached to the edge. Given that all the attachment points were occupied within 2 nm, this is not surprising. To spread the Si atoms out, the sub-scan area was moved back and forth along the edge as depicted by the box and arrows in f). The result, shown in the inset in f), is that the Si atoms have dispersed randomly. Continued sputtering of the source material, both above and below the edge, continued to attach Si atoms to the graphene. g) and h) show the result. We note that the imaging process itself is restructuring the edge and removing Si atoms, as evidenced by the broken and streaky appearance of the Si atoms in h), thus we were unable to achieve a fully passivated graphene edge where Si atoms occupied every attachment site.

### 2.2 Carbon Sculpting for Removal or Incorporation of Dopant Atoms

It is well known that the knock-on energy for graphene is about 80 kV.[35, 36] Thus, graphene is quite robust against the 60 kV beam used in these experiments. However, the graphene edge atoms are more susceptible to knock on damage from the beam.[34] In **Figure 2** a)-g) we illustrate how the beam restructures and sputters carbon edge atoms. The initial configuration is shown in a). d)-g) show a series of sequentially acquired images of a small scan area with many streaks appearing at the edge of the graphene. These result from atoms moving during the image acquisition and indicate the instability of the edge atoms. Because of sample drift, the images are not perfectly aligned so the same reference atom is circled in each image for comparison. We observe that about one layer of carbon edge atoms have been sputtered away during image acquisition but also a significant amount of rearrangement occurs without loss of atoms. The image in b) was acquired immediately afterword and the hole is slightly more pointed than it was in a). After about two minutes, the image in c) was acquired and the graphene edge has returned to a round shape. This is not surprising since it is expected that regions with the highest curvature will have the highest likelihood of capturing an atom or restructuring to lower the curvature.

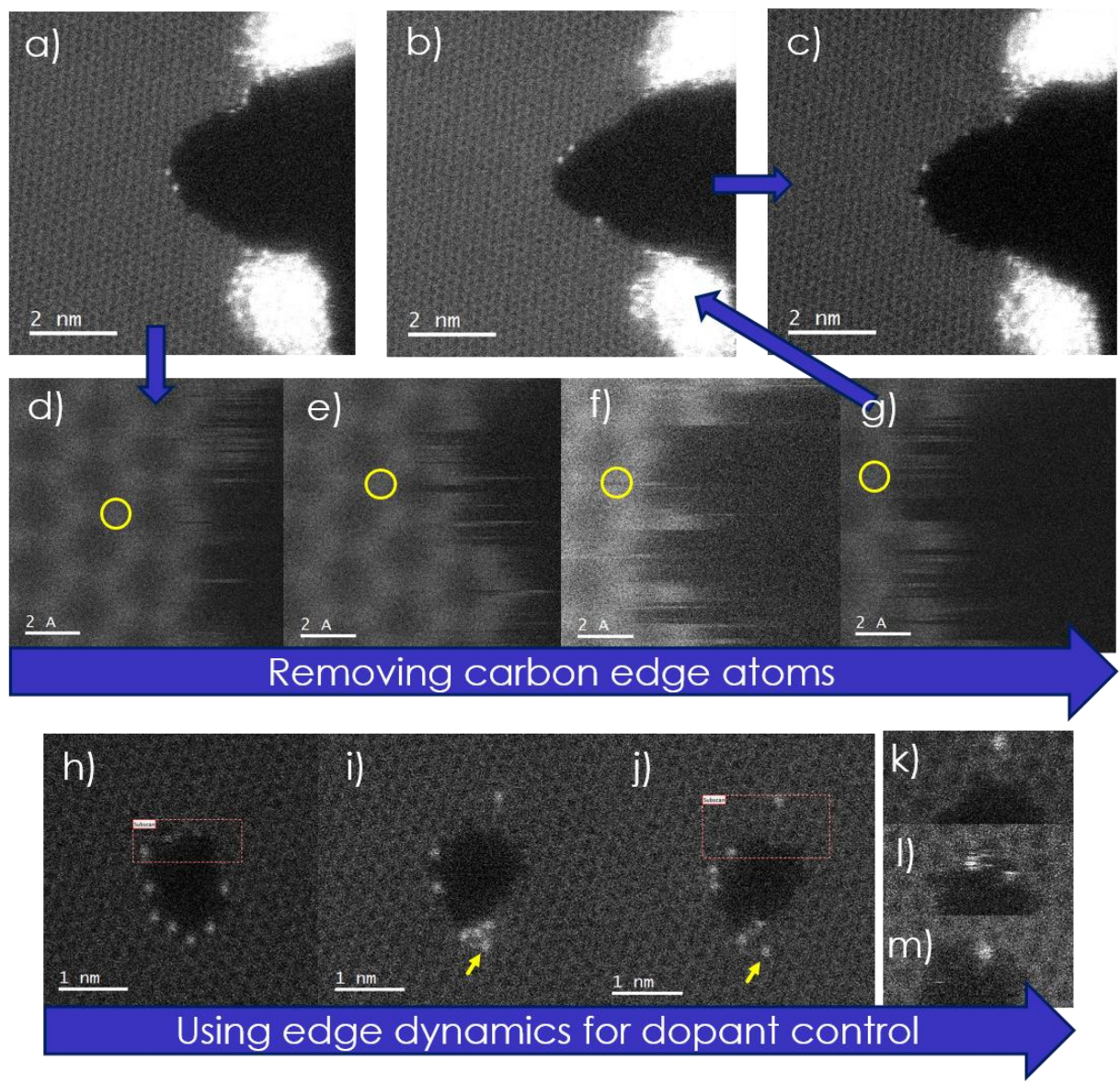

**Figure 2** a)-c) show the result of sputtering away carbon atoms from the edge. a) shows the initial configuration, b) shows the configuration immediately after, and c) shows the configuration two minutes later. Notably, the edge restructures from the pointed hole in b) to the round hole in c). d)-g) illustrate the process of sputtering away edge atoms by scanning the beam over a small area. Due to sample drift the images are not perfectly aligned, so the degree of sputtering appears larger than it is. The same atom is circled in each image as a reference point for the eye and we see that this image series documents the removal of only about one atomic layer of the edge rather than what initially appears to be two or three layers. h)-m) show how predictable edge dynamics may be harnessed for atomic-level control. h) shows the initial configuration of a small hole bearing a number of Si edge atoms. After performing a sub-scan over the top portion of the hole, indicated by the box in h), Si and C atoms cluster toward the bottom of the hole and the arrowed atom became buried in the graphene lattice, shown in i) and j). During the sub-scan process, a Si atom became stuck in the lattice above the hole, i) and j). Scanning this area again, resulted in growth (or movement) of the hole upward to recapture the Si atom within the hole. k)-m) are images acquired of the sub-scan area during the process.

Given the observations from Figure 1 and Figure 2 a)-g) we attempted to explore whether we can utilize the edge dynamics to "bury" the Si edge atoms in the lattice. In other words, since were unable to coax Si atoms away from the edge into the lattice through the C-Si bond inversion mechanism,[25] perhaps we can accomplish this by attaching C atoms to the edge on top of the Si atom. Figure 2 h)-j) shows the results of this experiment. A small hole in the graphene lattice was found with a number of Si atoms attached to the inner edge, h). A sub-scan area was defined (boxed in h)) and this portion of the lattice was exposed to electron irradiation for a few seconds. In the subsequent image, shown in i), we observe that this procedure is indeed "pushing" the Si atoms, on average, toward the bottom of the hole and one of them, indicated by the arrow, becomes buried. Continued scanning at the top of the hole allows for further restructuring and the arrowed Si atom becomes properly buried, shown in j), where it is no longer even adjacent to any other edge atoms. We suggest the driving force for the observed motion may be conceptualized as an atom mobility gradient introduced by the beam. Irradiated atoms are much more likely to move, thus, as they randomly move out of the beam path toward the lower portion of the hole, they become more likely to immobile. This procedure is tantamount to controllably producing directional hole or nanopore migration within the graphene lattice.

During the sub-scan process, a Si atom became serendipitously stuck in the graphene lattice at the top, i) and j). This immediately lent the opportunity to demonstrate the reverse process whereby the hole is drawn by the scanned beam toward the dopant atom and reincorporated as an edge atom. The sub-scan area used in this process is boxed in j) and subsequent images acquired during the sub-scan are shown in k)-m). Given these observations one can imagine a process to controllably introduce dopant atoms into the graphene lattice as follows:

1) create a small hole in the lattice near an atom source
2) sputter atoms from the source onto the edge of the graphene lattice
3) sputter C atoms from the opposite side of the hole to bury the foreign atoms and incorporate them into the lattice
4) (possibly) sputter C atoms from an amorphous C source material or from e-beam deposited C contamination to heal the hole.[11]

In attempts to master e-beam induced atomic control, additional tools must be developed to enable more sophisticated beam control. In the results presented here we have simply made use of a sub-scan box that can be moved around via mouse control within a larger reference image. This enables the microscopist to arbitrarily localize the beam and create controlled beam paths while *simultaneously* being able to observe a small image of the irradiated area. While this seems rather rudimentary, this simple tool is quite enabling. Imagine what could be done with a suite of such beam control tools. For example, it would be advantageous to develop a masking tool so that the beam is blanked automatically during image acquisition when it comes to a masked area. This would allow the operator to acquire images of the surrounding area while ensuring limited beam exposure to a more delicate feature of interest. Real-time noise removal and image reconstruction from sparse data would allow faster imaging and reduced unintentional beam exposure while imaging. Electronic drift compensation would also reduce the need for continuous exposure to the beam to locate atomic positions. Dynamic pixel dwell time may be useful to introduce controlled mobility gradients. Additionally, use of predefined triggers may prove to be quite powerful, where, for example, the hardware controller may be told to move the beam to a specified location for a certain amount of time or until the intensity on a detector crosses some threshold, for example. Enhancements of this nature would greatly improve the microscopist's ability to translate experimental ideas related to beam control into actual experiments.

## 3. Conclusions

In the pursuit of ultimate material control, moving and assembling single atoms and manipulating atomic scale structures is the foundation. The experiments described here illustrate ways to manipulate matter on sub-nanometer length scales using the finely focused probe of an aberration corrected STEM. We illustrated how Si atoms, found passivating the edges of graphene, could be removed through user/mouse-controlled beam movement. We further demonstrated that Si atoms could be sputtered from a nearby source material to re-passivate the graphene edge. Si atoms passivating the edge of a small hole could be incorporated into the graphene lattice by sputtering other edge atoms on top of them. One could think of this as gradually moving the hole away from the Si atom which leaves it stuck in the lattice. Alternatively, we showed how a Si substitutional atom stuck in the lattice near the hole could be reintroduced into the hole by moving the hole toward the defect (i.e. sputtering the edge of the hole). These simple demonstrations highlight the undiscovered possibilities for STEM to be used as a nanofabrication platform at the atomic scale. While still in its infancy, such ability would allow for the manufacture of exotic atomic configurations, study of atomic chemistry, and building of functional atomic machines. While these experiments were performed using Si, they likely extend to other atoms as well. Ramassi et. al.,[13] for example, have shown Ti, Ni, Al, and Pd also attach to the edges of graphene. Finally, we commented on what we believe to be a dearth of practical beam control tools available to the microscopist and make some suggestions for future enhancements which would enable more carefully crafted experiments and easier microscope control.

## 4. Experimental

CVD-grown graphene was transferred from the Cu foil growth substrate to a TEM sample grid followed by an $Ar/O_2$ anneal for removal of volatile adsorbents. The Cu foil was spin-coated with poly(methyl methacrylate) (PMMA) to stabilize the graphene and the Cu foil was etched away in a bath of ammonium persulfate-deionized (DI) water solution. The graphene/PMMA layer was transferred to a DI water bath to remove residues of ammonium persulfate. The graphene was transferred to the TEM substrate by scooping it from the bath and letting it dry at room temperature. To produce better adhesion to the TEM substrate, the sample was heated at 150 C on a hot plate, followed by an acetone bath heated at 80 C for 15 min to remove PMMA. Samples were then rinsed in isopropyl alcohol and baked in an oven under an $Ar/O_2$ (90%/10%) environment to remove residual PMMA and volatile organic compounds.[37, 38]

Electron beam experiments were performed in a Nion UltraSTEM U100 at an accelerating voltage or 60 kV in high angle annular dark field (HAADF) imaging mode. The samples were loaded into the microscope using our standard loading procedure, where the microscope magazine, cartridges, and samples are baked in a vacuum chamber at 160 °C for eight hours prior to insertion into the microscope.

**Acknowledgements**

We would like to thank Dr. Ivan Vlassiouk for provision of the graphene samples and Dr. Francois Amet for assisting with the argon-oxygen cleaning procedure.

Research supported by Oak Ridge National Laboratory's Center for Nanophase Materials Sciences (CNMS), which is sponsored by the Scientific User Facilities Division,

**Table of contents entry:**
**Current investigations in atomic-level material manipulation in the scanning transmission electron microscope appear promising.** In this article, the use of controlled e-beam dosing is explored to detach and reattach Si atoms passivating graphene edges. In addition, controlled e-beam dosing is used to "bury" Si atoms in the graphene lattice or "excavate" them from the lattice.

**Keyword** electron beam manipulation, graphene, scanning transmission electron microscope, dopant control

**Authors** O. D. Author 1 Corresponding Author, S. K. Author 2, S. V. K. Author 3, S. J. Author 4

**Title** E-beam manipulation of Si atoms on graphene edges with aberration-corrected STEM

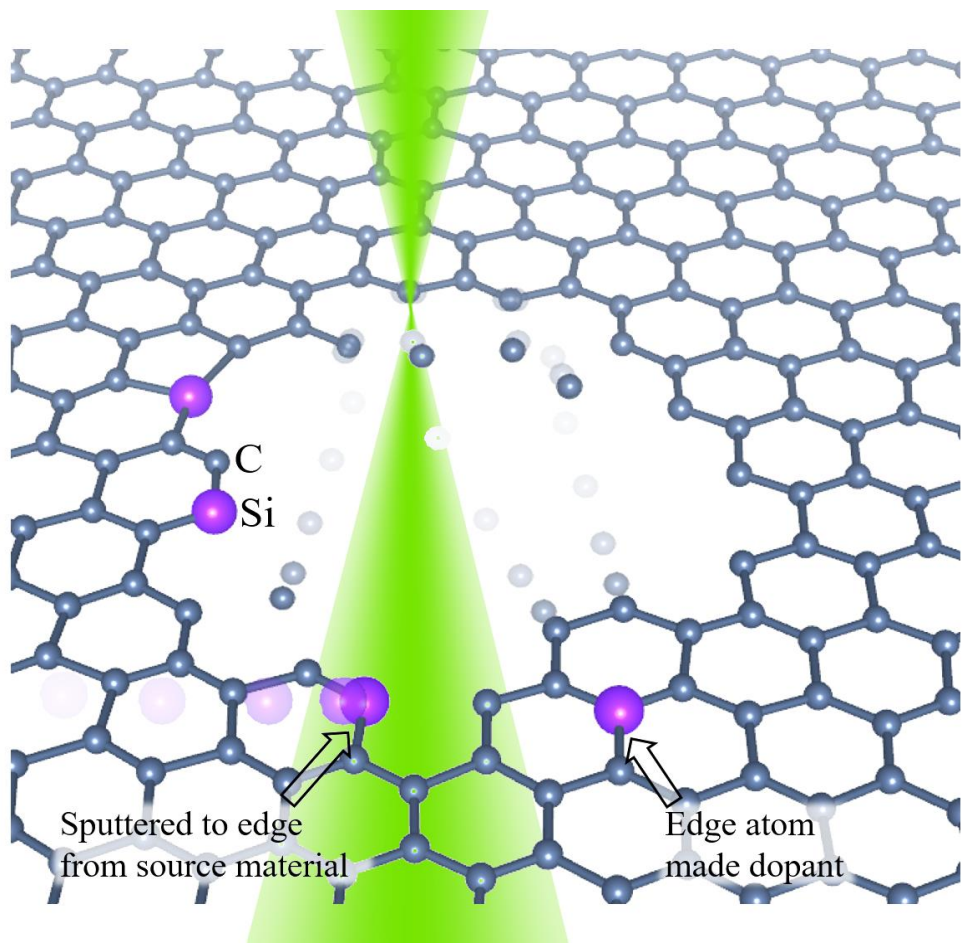